# Modeling of a piezoelectric micro-scanner

A. Chaehoi∗, M. Begbie∗, D. Cornez•, K. Kirk•

∗ Institute for System Level Integration, Alba Center, Alba Campus, EH547EG Livingston, Scotland, www.isli.co.uk

• University of Paisley School of Engineering and Science, Paisley Campus, Paisley PA1 2BE, Scotland, www.paisley.ac.uk

Contact: Aboubacar Chaehoi, tel: +44 1506 469 307, fax: +44 1506 469 301, aboubacar.chaehoi@sli-institute.ac.uk

**Abstract**

Micro-scanners have been widely used in many optical applications. The micro-scanner presented in this paper uses multimorph-type bending actuators to tilt a square plate mirror. This paper presents a complete analytical model of the piezoelectric micro-scanner. This theoretical model based on strength of material equations calculates the force generated by the multimorphs on the mirror, the profile of the structure and the angular deflection of the mirror. The proposed model, used to optimize the design of the piezoelectric silicon micro-scanner, is intended for further HDL integration, allowing in this way system level simulation and optimization.

## I. Introduction

An optical micro-scanner is usually composed of an adjustable mirror that is used to reflect light beams. Micro-scanning mirrors have a large variety of applications such as optical scanning [1], display devices [2], printers or barcode scanning. Many actuating principles for the adjustment of the mirror can be found. For instance the mirror plate can be actuated electrostatically; the driving electrodes can be a layer under the mirror itself or comb fingers linked with it.

The Laplace force can be used to magnetically actuate the mirror. Inductances are placed one on the mirror, another on the substrate, the deflection of the mirror is obtained when a magnetic field is created by applying a current through the coils.

Thermally actuated micro-scanners use bi-layer thermal bimorph beams. Two active layers compose the beam. Due to different thermal expansion coefficients or different geometries, variation of temperature causes one layer to expand more than the other, leading to a bending of the beam and the attached mirror.

Piezoelectric bimorphs can also be used to actuate the mirror. In this case, an electric field causes one layer to extend and the other layer to contract. Several works have been published on modeling of bimorphs or multimorphs, as a device is named when it consists of more than two layers. Smits and Choi [3] presented the constituent equations of piezoelectric heterogeneous bimorphs, deriving the static behaviour of bimorph structures from basic thermodynamic principles.

Based on the Timoshenko's approach for the deflection of a thermal bimorphs, Chu et al. [4] presented a model of a cantilever piezoelectric bimorph. Huang, Lin and Tang [5] also developed a new method to obtain the analytic expression of the curvature and the displacement of a multilayer piezoelectric cantilever.

In this paper we develop a complete static model of a micro-scanner that includes two piezoelectric multimorphs and an attached mirror. The analytical expression of the profile all along the beams and mirror will be presented. From this profile, the deflection angle of the mirror, the primary parameter of the micro-scanne,r will be deduced.

## II. The piezoelectric micro-scanner

This study concerns the modeling of a simple micro-scanner; as depicted in Figure 1. A rigid mirror plate is attached to two piezoelectric multimorphs. The multimorphs are composed of two piezoelectric layers deposited over a silicon substrate. When one bimorph is driven by a voltage, the beam will bend in one direction. Supplying the other bimorph with an opposite voltage makes that beam bend in the opposite direction. Doing so we obtain an angular deflection angle of the mirror about a central axis that is fixed in the z-direction.

The micro-scanner can be modeled as three attached beams loaded by two opposite forces (Fig. 1). Beam [BC] represents the mirror with a length $L_2$, a Young modulus $E_2$ and a moment of inertia $I_2$. Beams [AB] and [CD] are fully constrained at A and D, they represent the actuating piezoelectric beams with length $L_1$, Young modulus $E_1$ and moment of inertia $I_1$. The opposing loads F are applied at B and C, they represent the force exerted on the mirror by the piezoelectric bimorphs.

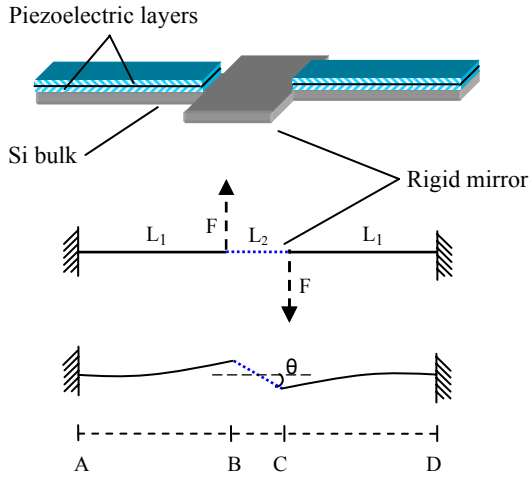

Figure 1: The piezoelectric micro-scanner
showing the bimorph–rigid plate–bimorph construction

We want to calculate the deflection of the beams and the tilt of the mirror plate. The bending of a beam can be obtained from the bending moment [6] as

$$E.I \frac{d^2 y(x)}{dx^2} = M(x) \tag{1}$$

$$y(x) = \iint \frac{M(x)}{E.I} dx.dx \tag{2}$$

where E is the Young modulus, I is the moment of inertia, $y(x)$ is the bending and $M(x)$ is the bending moment.

Since the three beams don't have the same mechanical properties we can't use the superposition theorem for the calculation of the bending moment along the whole beam. However the model can be simplified. Since beams [AB] and [CD] have the same properties, the center of the mirror is then a fixed point and we have a symetry in the problem. The simplified model is depicted in Figure 2, one end of the beam is fully constrained while the other end (the mirror center) is on a support allowing free rotation about one axis. $E_1$ is the Young modulus of the beam [AB], $I_1$ is the moment of inertia of the beam, $E_2$ $I_2$ are the material properties of the equivalent homogeneous bimorph beam.

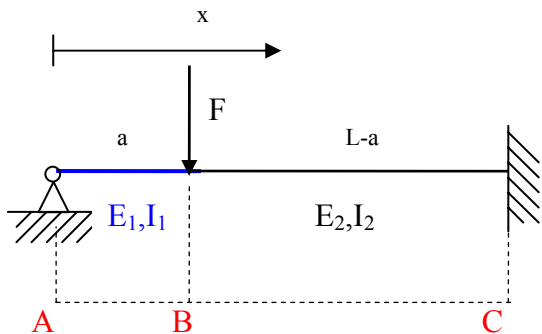

Figure 2: Simplification by symetry of the problem

We can write the folowing equations along [AB]:

$$T_{[AB]}(x) = -R_A \tag{3}$$

$$M_{[AB]}(x) = R_A.x \tag{4}$$

Where T is the shearing stress, $R_A$ is the reaction of the support at A and M is the bending moment. Along [BC] we can write:

$$T_{[BC]}(x) = F - R_A \tag{5}$$

$$M_{[BC]}(x) = (F + R_A)x - F.a \tag{6}$$

Since the moment of inertia of the bimorph beams is low compared to that of the mirror, we can consider that the mirror is rigid. With this assumption, the curvature along [AB] is nil, $\frac{M(x)}{E_1 I_1} = 0$.

At the fixed end (C) we have $y(x=L)=0$ and $dy(x=L)=0$.

At the support (A) we have $y(x=0)=0$.

And at the junction of the two beams (B) we have two continuity conditions on $y(x)$ and $dy(x)$.

With all six conditions and by integrating the expression of the bending moment twice (equation 1), we obtain the following expressions (the solution can be more easily computed using Matlab®) :

$$R_A = \frac{-F(a^3 - 3aL^2 + 2L^3)}{2L^3 - 2a^3} \tag{7}$$

$$y_{[AB]}(x) = \frac{-Fa(a-L)^3 x}{4E_2 I_2 (a^2 + La + L^2)} \tag{8}$$

$$y_{[BC]}(x) = \frac{Fa[(a+L)x^3 + x^2(-2L^2 - 2a^2 - 2aL) + x(L^3 + 4a^2 L + aL^2) - 2a^2 L^2]}{4E_2 I_2 (a^2 + La + L^2)} \tag{9}$$

We obtain then:

$$\phi = \arctan\left(\frac{Fa(L-a)^3}{4E_2 I_2 (a^2 + La + L^2)}\right) \tag{10}$$

where $\Phi$ is the tilt of the mirror.

### III. Model of the multimorph force

In the previous section we have presented a model of the device using the force F generated by the piezoresitive bimorph as parameter. The following section concerns the calculation of this force.

Smits and Choi [3] have presented the constituent equations of piezoelectric heterogeneous bimorphs. Those equations, derived from thermodynamic principles, describe the static and dynamic behaviour of bimorphs structures. Huang, Lin and Tang [5] have developed a new method to obtain the analytical expression of the curvature and the displacement of a multilayer piezoelectric cantilever. We have used this method to get the equivalent force generated by our three layers bimorphs.

Figure 3 is a cross section of the three layer piezoelectric bimorphs. $R$ is the radius of the curvature, $t_s$ and $t_p$ are the thickness of the substrate and the piezoelectric layers. For a given layer $i$, $P_i$ is the average stress while $M_i$ is the average moment of force in the x-y plane. $E$ is the Young modulus and $S$ is the strain in the layers due to the piezoelectric action. Here the electrodes are not represented; we neglect their effects due to their thickness.

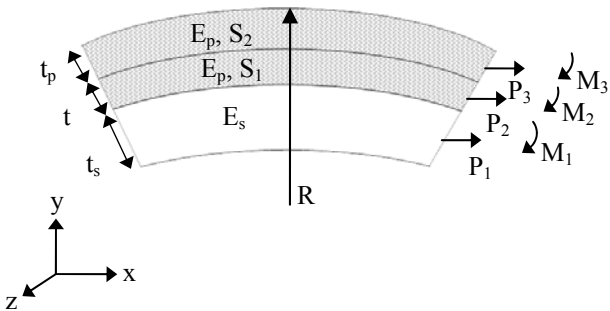

Figure 3: Cross section of a 3 layer piezoelectric multimorph

The curvature $1/R$ of the curve $y(x)$ is expressed as follows:

$$\frac{1}{R} = \frac{d^2 y}{dx^2} \quad (11)$$

then

$$y(x) = \frac{x^2}{2R} \quad (12)$$

The tip deflection is found at $x=L$. In the following we are going to determine $1/R$ in order to get the tip deflection of the beam.

According to Figure 3 and to satisfy stress equilibrium in the x-y plane

$$P_1 + P_2 + P_3 = 0 \quad (13)$$

The moment equilibrium gives

$$\frac{t_s}{2} P_1 + \left(t_s + \frac{1}{2} t_p\right) P_2 + \left(t_s + \frac{3}{2} t_p\right) P_2 + \frac{E_s t_s^3 + 2 E_p t_p^3}{12 R} = 0 \quad (14)$$

The junction between the substrate and the bottom piezoelectric layer must satisfy the interface continuity condition (the strain is present only in the piezoelectric layer)

$$\frac{P_1}{E_s t_s} + \frac{t_s}{2R} = \frac{P_2}{E_p t_p} - \frac{t_p}{2R} + S_1 \quad (15)$$

From the interface continuity condition of the two piezoelectric layers, we can write

$$S_1 + \frac{P_2}{E_p t_p} + \frac{t_p}{2R} = \frac{P_3}{E_p t_p} - \frac{t_p}{2R} + S_2 \quad (16)$$

We can put this system of equations in a matrix form, we get

$$\begin{bmatrix} 1 & 1 & 1 & 0 \\ \frac{t_s}{2} & t_s + \frac{t_p}{2} & t_s + \frac{3t_p}{2} & \frac{E_s t_s^3 E_p t_p^3}{E_p t_p} \\ \frac{1}{E_s t_s} & -\frac{1}{E_p t_p} & 0 & \frac{t_s + t_p}{2} \\ 0 & \frac{1}{E_p t_p} & -\frac{1}{E_p t_p} & tp \end{bmatrix} * \begin{bmatrix} P_1 \\ P_2 \\ P_3 \\ \frac{1}{R} \end{bmatrix} = \begin{bmatrix} 0 \\ 0 \\ S_1 \\ S_2 - S_1 \end{bmatrix} \quad (17)$$

The strains in the piezoelectric layers are expressed as:

$$S_1 = -\frac{d_{31}}{t_p} V \quad (18)$$

$$S_2 = \frac{d_{31}}{t_p} V \quad (19)$$

where $d_{31}$ is the piezoelectric coefficient relating piezoelectric field to strain according to the x-y plane and $V$ is the voltage apply across each piezoelectric layer.

Extracting $\frac{1}{R}$ from equation 17 and using equations 18, 19 and 12 we get the tip deflection $y_L$

$$y_L = 6 \frac{L^2 E_p t_p d_{31} (E_s t_s + 2 E_p t_p)}{8 E_s t_s^3 E_p t_p + 24 E_s t_s^2 E_p t_p^2 + 32 E_s t_s E_p t_p^3 + E_s^2 t_s^4 + 16 E_p^2 t_p^4} V \quad (20)$$

The equivalent external force that will lead to the same displacement is obtained by multiplying the tip deflection by the stiffness of the whole beam.

$$F = K \cdot y_L = \frac{3 E_{eq} I_{eq}}{L^3} y_L \quad (21)$$

Where $K$ is the stiffness, $E_{eq}$ and $I_{eq}$ are respectively the equivalent Young modulus and moment of inertia of the 3-layers beam.

The next section concerns the determination of the equivalent beam that has the same mechanical properties $E.I$ and elastic behaviour as the 3-layer multimorph.

### IV. Equivalent section

The method to calculate the equivalent beam is described in Figure 4. This method is based on the construction of an equivalent homogeneous section that is mechanically equivalent to the initial section. For that, the width $W_i$ of each layers in the beam is reduced by the ratio of its Young modulus over the highest Young modulus present in the beam (method of normalization of the widths by the elasticity modulus):

$$W_{i\ rectified} = W \frac{E_i}{E_{max}} \quad (22)$$

This method guarantees the conservation of the flexural rigidity which corresponds to the product $E.I$. In our case the highest Young modulus correspond to the piezoelectric layer, $E_{max} = E_3$.

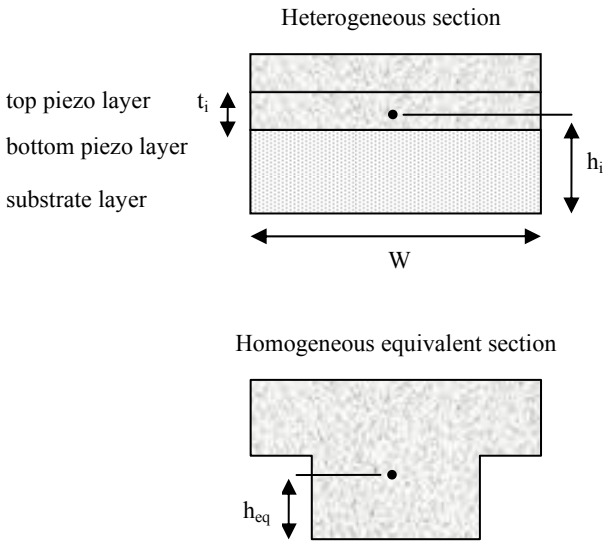

Figure 4: a) A heterogeneous beam of width W, with overlying piezoelectric layers of thickness $t_i$ and with their neutral axes at a height $H_i$. b) The equivalent homogeneous beam section.

In the homogeneous section the neutral axis $h_{eq}$ is located at the barycentre of the stiffnesses. It corresponds to the barycentre of the surfaces $S_i = t_i.W_{i\ corrected}$.

In our case the piezoelectric layers have the same thickness, so $t_2=t_3$. It comes

$$h_{eq} = \frac{\sum S_i.h_i}{\sum S_i} = \frac{t_1^2 E_1 + 4t_3^2 E_3 + 4t_1 t_3 E_3}{2(t_1 E_1 + 2t_3 E_3)} \quad (23)$$

The equivalent moment of inertia is expressed as follows

$$I_{eq} = W \left( \sum \frac{E_i \left( \frac{t_i^3}{12} + t_i (h_{eq} - h_i)^2 \right)}{E_{max}} \right)$$

$$I_{eq} = \frac{W \left( 8E_1^3 t_3^3 E_3 t_3 + 24 E_1^2 t_1^2 E_3^2 t_3^2 + 32 E_1 t_1 E_3^3 t_3^3 + E_1^2 t_1^4 + 16 E_3^2 t_3^4 \right)}{E_3 (t_1 E_1 + 2t_3 E_3)}$$
(24)

From equations 24, 20 and 21 we can calculate the equivalent force generated by the piezoelectric bimorph on the beam.

Assuming that the substrate is the first layer and the piezoelectric layers are layer$_2$ and layer$_3$, $E_1 = E_s$, $E_3 = E_p$, $t_1 = t_s$ and $t_3 = t_p$, it comes

$$F = \frac{3}{2} \frac{W_p t_p E_p d_{31}}{L} V = \frac{3}{2} \frac{W_p t_p d_{31}}{L.S_{11}^E} V \quad (25)$$

Where $S_{11}^E$ is the compliance coefficient at constant electric field.

### V. Micro-scanner modeling results

Different sizes of the structure depicted in Figure 1 have been modelled. The substrate we have chosen is silicon while the piezoelectric material is PZT-5H. The central plate mirror is 300μm square, the thickness of the substrate layer is 5μm and the thickness of each piezoelectric layer is 1μm. The profile of the micro-scanner under 50V applied voltage is depicted in Figure 5. The tilt angle of the mirror $\Phi$ and the maximum deflection of the beams $y_{max}$ obtained from equations 8, 9 and equation 25, are given in Table 1. From this model the profile of the structure is also calculated. In figure 5, we can see the profile of different micro-scanners when they are driven with 50V.

| Micro-scanner | A | B | C |
|---|---|---|---|
| beam size (μm) | 850*30 | 600*30 | 500*30 |
| $\Phi$ (deg) | 0.57 | 0.48 | 0.42 |
| $y_{max}$ (μm) | 2.45 | 1.76 | 1.48 |

Table 1: Tilt and maximum deflection of the device under 50V.

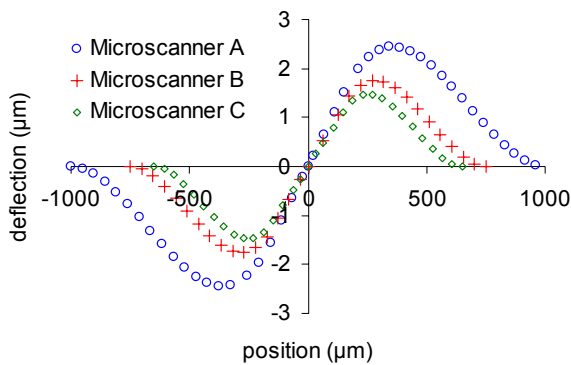

Figure 5: Profile of the micro-scanner under 50V.

Large angular deflection of the mirror is obtained with long beams. From Figure 5 we can see that an angle around 0.6 degrees can be obtained with this very simple structure.

### VI. Conclusion

This paper concerns the modeling of a piezoelectric micro-scanner. A simple structure for a micro-scanner has been proposed. The device is composed of two three-layer bimorph beams that are used to actuate a mirror plate. We have proposed a complete analytic model of the piezoelectric device that leads to the profile of the structure and the tilt of the mirror. This model can be used very rapidly to study the effects of geometrical parameters and material properties on the performance of the micro-scanner. The proposed model will be used as a basis for further behavioural modeling and VHDL system integration.

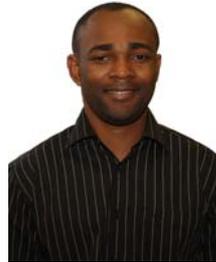

Aboubacar Chaehoi received his Ph.D. degree in Micro-electronics from Montpellier University in 2005. From 2001 he worked on micro-electromechanical systems design and modeling at the Montpellier Laboratory of Computer Science, Robotics, and Microelectronics (LIRMM - France). His work concerned the design and characterization of a 3-axis accelerometer in CMOS technology (using both piezoresistive and thermal transduction) and the development of FEM and behavioral models for those sensors. He is currently a MEMS designer at the Institute of System Level Integration (ISLI) in Scotland, where he is involved in design of MEMS for optical and micro-photonic applications.